\documentclass[10pt,aps,prl,twocolumn,superscriptaddress]{revtex4-1}
\usepackage{graphicx}
\usepackage{subfig}
\usepackage{floatrow}
\usepackage{amsmath}
\usepackage{hyperref}
\usepackage{floatrow}
\usepackage{siunitx}
\usepackage {ulem}
\usepackage{caption}

\usepackage{ragged2e}
\usepackage{dcolumn}
\usepackage{bm}

\begin{document}

\preprint{APS/123-QED}

\title{Non-ohmic to ohmic crossover in the breakdown of the quantum Hall states in graphene under broadband excitations}

\author{Torsten Röper}
\email{roeper@ph2.uni-koeln.de}
\affiliation{II. Physikalisches Institut, Universit\"at zu K\"oln, Z\"ulpicher Str. 77, D-50937 K\"oln, Germany}
\author{Aifei Zhang}
\affiliation{Université Paris-Saclay, CEA, CNRS, SPEC, 91191 Gif-sur-Yvette, France}
\author{Kenji Watanabe}
\affiliation{Research Center for Electronic and Optical Materials,
National Institute for Materials Science, 1-1 Namiki, Tsukuba 305-0044, Japan}
\author{Takashi Taniguchi}
\affiliation{Research Center for Materials Nanoarchitectonics,
National Institute for Materials Science, 1-1 Namiki, Tsukuba 305-0044, Japan}
\author{Olivier Maillet}
\affiliation{Université Paris-Saclay, CEA, CNRS, SPEC, 91191 Gif-sur-Yvette, France}
\author{François D. Parmentier}
\affiliation{Université Paris-Saclay, CEA, CNRS, SPEC, 91191 Gif-sur-Yvette, France}
\affiliation{Laboratoire de Physique de l’Ecole Normale Supérieure, ENS, Université PSL, CNRS, Sorbonne Université, Université Paris Cité, F-75005 Paris, France}
\author{Erwann Bocquillon}
\affiliation{II. Physikalisches Institut, Universit\"at zu K\"oln, Z\"ulpicher Str. 77, D-50937 K\"oln, Germany}

\date{\today}

\begin{abstract}
Graphene, through the coexistence of large cyclotron gaps and small spin and valley gaps, offers the possibility to study the breakdown of the quantum Hall effect across a wide range of energy scales. In this work, we investigate the breakdown of the QHE in high-mobility graphene Corbino devices under broadband excitation ranging from DC up to \SI{10}{\giga\hertz}. We find that the conductance is consistently described by variable range hopping (VRH) and extract the hopping energies from both temperature- and field-driven measurements. Using VRH thermometry, we are able to distinguish between a cold and hot electron regime, which are dominated by non-ohmic VRH and Joule heating, respectively. Our results demonstrate that the breakdown in the quantum Hall regime of graphene is governed by a crossover from non-ohmic, field-driven
VRH to ohmic, Joule-heating–dominated transport.

\end{abstract}

\maketitle


    \paragraph{\label{sec:Intro} Introduction --} 
    
     In the Quantum Hall (QH) regime, quantization arises from bulk-localized states induced by disorder, which protect the edge channels from backscattering. However, sufficiently strong perturbations—such as elevated temperatures or large electric fields—can activate bulk conduction and destroy quantization. Understanding the breakdown mechanisms of the QH effect is essential for both fundamental studies of localization and practical applications involving dissipationless transport, such as quantum metrology \cite{ribeiro2015,fijalkowski2024}. The breakdown process is typically described by variable range hopping (VRH) transport~\cite{efros_1975, Shklovskii2024, Polyakov1993} and has been studied in DC and low-frequency regimes in graphene~\cite{ Hohls2002,Bennaceur2012,Baker2012} and other two-dimensional electron systems~\cite {Furlan1998,Ladieu2000}. Systematic investigations at radio and microwave frequencies remain scarce. They are nonetheless particularly relevant for high-frequency transport in edge states \cite{Dartiailh2020, bocquillon2013b, hashisaka2017,Kamata2022,Gourmelon2023}, the manipulation of single electron states in electron quantum optics experiments \cite{bocquillon2013,bocquillon2014}, or for proposed applications—including non-reciprocal microwave components— which operate in the GHz regime~\cite{viola_hall_2014,mahoney_zero-field_2017,roeper2024}.
    
    In graphene, the unique fourfold spin and valley degeneracy of the Landau levels leads to a hierarchy of energy gaps: the robust cyclotron gaps and the much smaller symmetry-broken gaps due to spin or valley polarization,  spanning energy gaps from a few kelvins to several hundreds of kelvins at \SI{9}{\tesla} \cite{Young2012}. Recent studies have shown that the characteristic hopping energy $T_0$, extracted from temperature-dependent bulk conductance measurements, varies significantly across filling factors~\cite{Aifei2025,Kaur2024}.

    \begin{figure}[t]
        \centering
        \includegraphics[width=0.85\textwidth]{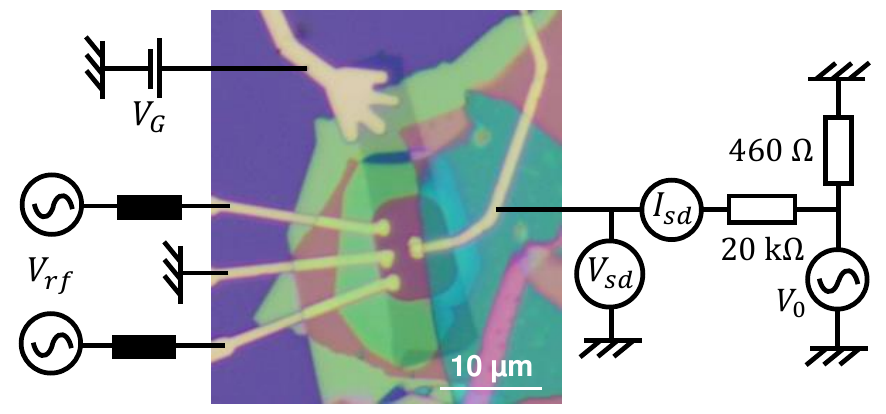}
        \caption{\justifying\textbf{Sample layout and experimental setup:} Microscope image of the device. The top contact is connected to a graphite back gate (grey contrast). The graphene Corbino ring is sandwiched between hBN flakes (shown in green). The blue flake is an additional hBN flake, avoiding shorts between the graphene and the central ohmic contact (right). The left ohmic contacts are grounded, and two are connected to RF sources via bias-tees. The conductance is obtained by voltage-biasing the central contact.}
        \label{fig:device_setup}
    \end{figure}
    
    \begin{figure*}[t]
        \centering
        \includegraphics[width=0.9\textwidth]{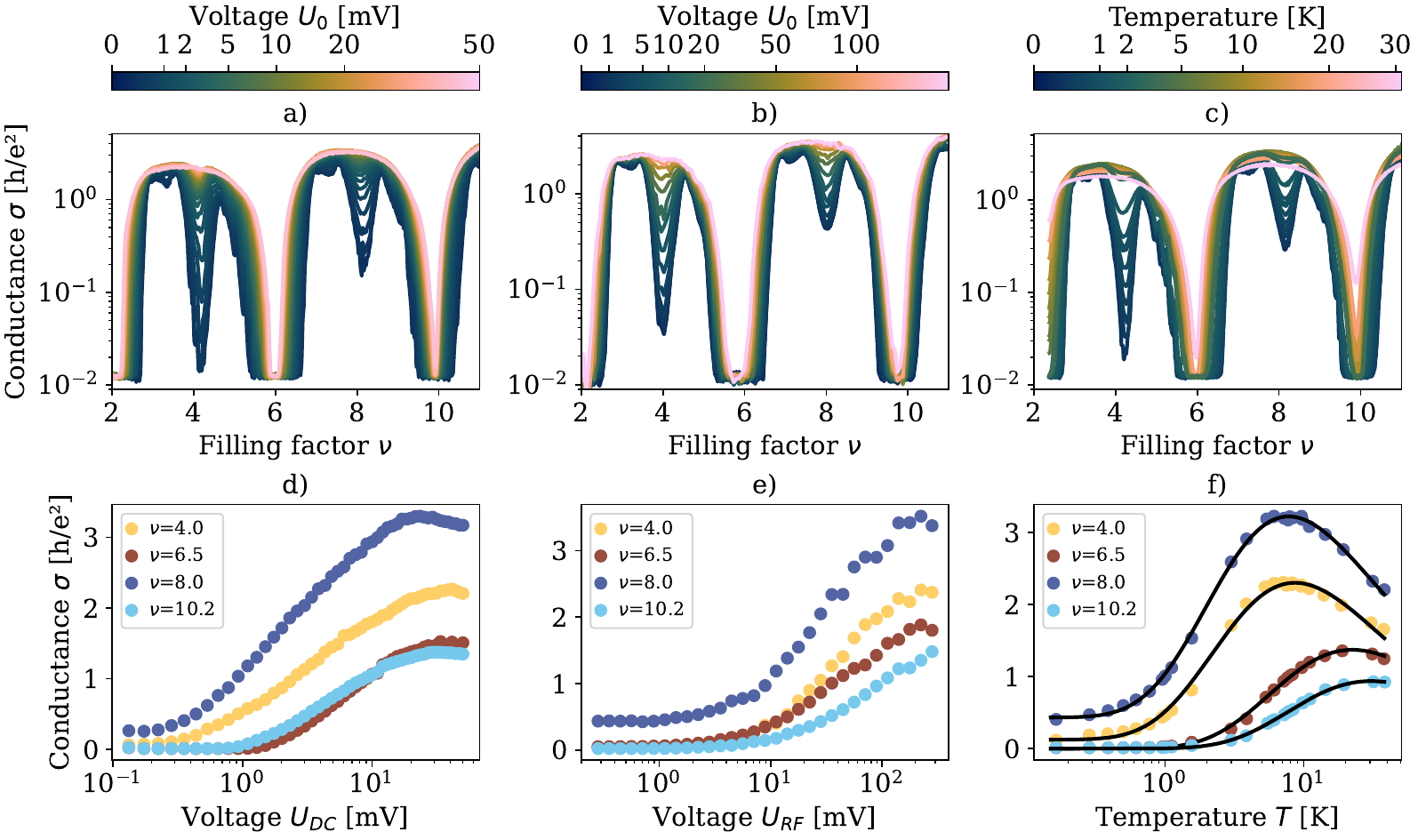}
        \caption{\justifying\textbf{Comparison of breakdown mechanisms under DC/RF excitation and temperature.}
        (a–c) Conductance vs. filling factor $\nu$ for increasing DC bias voltages (a), increasing microwave power at \SI{6}{\giga\hertz} (b), and increasing temperature (c).
        (d–f) Conductance vs. DC bias voltage $U_{\rm DC}$ (d), microwave amplitude $U_{\rm RF}$ (e), and temperature $T$ (f) for different filling factors. The black lines in f) correspond to a fit with Eq.~\ref{eg:VRH_temp} with $T_0$ and $\sigma_0$ as fitting parameters (with an offset for not fully developed gaps).}
        \label{fig:results}
    \end{figure*}

    Here, we report on the breakdown of QH states in graphene under DC and AC radiation up to \SI{10}{\giga\hertz}. An excitation signal is applied while measuring the DC conductance, providing an independent probe of the bulk response. For comparison, the breakdown of the quantum anomalous Hall effect studied with the same technique was dominated by Joule heating from microwave power dissipation, with a threshold that decreased systematically with frequency, attributed to enhanced bulk conduction via charge-puddle networks \cite{roeper2025}. This work examines whether analogous mechanisms govern breakdown in the QH regime of graphene.

    We analyze the applicability of the VRH framework to QH breakdown in high-mobility graphene Corbino devices. We extract the effective electronic temperature $T_{\text{eff}}$ via inversion of $\sigma(T)$ calibration curves and study its dependence on the applied bias. We find that  $T_{\text{eff}}$ follows a power-law behavior $T_{\text{eff}}\propto U^\beta$, where the exponent $\beta$ reflects the underlying transport mechanism. \\
    By tuning the filling factor, we access a broad range of hopping energies and can systematically track the breakdown mechanism across different localization strengths. This broad coverage enables us to resolve the crossover from non-ohmic to ohmic transport, overlooked in earlier studies~\cite{Bennaceur2012,Baker2012,Furlan1998,Ladieu2000} that primarily focused on large cyclotron gaps and thus probed only the high–$T_0$ regime. Our key finding is that QH breakdown is not governed by a single mechanism, but by a universal crossover between field-driven VRH and heating-dominated transport, controlled solely by the localization energy $k_B T_0$. This establishes the localization strength $\xi$ as the key parameter governing the onset of heating and dissipation in QH breakdown \cite{Aifei2025}.

    \paragraph{\label{sec:Device} Experimental Setup --}
    To investigate bulk transport in monolayer graphene, the material is patterned into a Corbino ring geometry (see Fig.~\ref{fig:device_setup}). The graphene is encapsulated between hBN flakes, ensuring high quality, suppressing charge inhomogeneities, and stabilizing symmetry-broken QH states, while the bottom hBN layer (\SI{50}{\nano\meter}) serves as a dielectric for the graphite back gate that tunes the filling factor $\nu$.

    The Corbino disk has one inner ohmic contact (right) and three grounded outer contacts (left), two of which are connected to microwave sources via bias-tees. The device is slightly anisotropic, with lateral dimensions of $\num{7.5}\times\SI{10}{\micro\meter^2}$ and a distance of $w\approx\SI{2}{\micro\meter}$ between the inner and outer contact.
    Measurements are performed in a dilution refrigerator at approximately \SI{20}{\milli\kelvin}, under a perpendicular magnetic field of \SI{9}{\tesla}, where the QH regime is fully developed. Contact resistances are confirmed to be negligible via two-terminal conductance tests in the metallic regime.

    To probe the bulk conductance, a \SI{100}{\micro\volt} AC voltage at \SI{7}{\hertz} is applied to the inner contact, and both current and voltage are recorded using standard lock-in techniques. The DC breakdown is characterized by applying a DC offset $U_{\rm DC}$. For microwave breakdown, continuous-wave RF signals of amplitude $U_{\rm RF}$ are applied to the outer contacts. Breakdown measurements from DC to \SI{10}{\kilo\hertz} use the lock-in source, while those between \SI{10}{\kilo\hertz} and \SI{10}{\giga\hertz} use a VNA. While the absolute amplitude of the high-frequency excitation (above \SI{1}{\giga\hertz}) carries a systematic uncertainty due to the difficulty in calibrating high-frequency signals, the relative frequency dependence and the trend across different filling factors remain reliable.\\
    This work focuses on measurements at high magnetic field ($B=\SI{9}{\tesla}$) and electron hoping ($\nu>0$), but we show in the SOM~\cite{supplement} that our claims also hold for smaller field and hole doping ($\nu<0$). The blue curves in Fig.~\ref{fig:results}a)-c) show the bulk conductance at low bias and temperature as a function of the filling factor $\nu$. The cyclotron gaps ($\nu=\num{2},\num{6},\num{10}$) and the spin gap at $\nu=\num{4}$ are fully developed, while the valley gaps ($\nu=\num{3},\num{5},\num{7}$) and the spin gap at $\nu=\num{8}$ show a residual conductance, because their gaps are not fully developed at $B=\SI{9}{\tesla}$, which is the maximum available magnetic field in our experiment. In the following, we will primarily investigate the cyclotron gaps and spin gaps as we examine VRH, and the valley states exhibit nearly metallic behavior.

    \paragraph{\label{sec:breakdown} Observation of the electric field driven breakdown --}
    First, we examine the breakdown under a constant voltage bias. The corresponding conductance $\sigma$ as a function of the filling factor $\nu$ is shown in Fig.~\ref{fig:results}a) for bias voltages $U_{\rm DC}$ ranging from 0 (blue) to \SI{50}{\milli\volt} (green). We observe that the spin gaps ($\nu=\num{4},\num{8}$) are vanishing already below $U_{\rm DC}\simeq\SI{1}{\milli\volt}$, as expected due to their relatively small size compared to the cyclotron gaps. These stay insulating until the maximum bias voltage of \SI{50}{\milli\volt} when the Fermi level lies in the center between two Landau levels, but becomes narrower with increasing bias voltage.\\  
    The breakdown under a microwave drive $U_{\rm RF}$ shows the same phenomenology, as shown in Fig.~\ref{fig:results}b) for a frequency of \SI{6}{\giga\hertz}. The only difference lies in an overall lower breakdown voltage, which we attribute to uncalibrated losses in the cable. However, they exhibit the same relative voltage dependence. Moreover, measurements over a wide frequency range from DC to \SI{10}{\giga\hertz} (Fig.~\ref{fig:frequency}) confirm that the shape and voltage scaling of the breakdown curves remain unchanged, further supporting that the underlying mechanism is frequency-independent.\\
    To further investigate the microscopic origin of this breakdown mechanism, we next examine its temperature dependence to identify the relevant energy scales and clarify whether both share a common VRH origin.
    
    \paragraph{\label{sec:temp}Observation of the temperature driven breakdown --}
    Fig.~\ref{fig:results}c) shows curves of the conductance $\sigma$ as a function of the filling factor $\nu$ for increasing temperatures $T$. The curve shows qualitatively the same phenomenology as the electric-field driven breakdown in DC and RF: The spin gaps ($\nu=\num{4},\num{8}$) are vanishing around \SI{1}{\kelvin}, while the cyclotron gaps only vanish at tens of Kelvin. The corresponding conductance as a function of the temperature $T$ is shown in Fig.~\ref{fig:results}f) for 4 different filling factors as in Fig.~\ref{fig:results}d)-e). We observe a similar shape as for the electric field-driven breakdown, but the range of accessible energies is larger than that accessible with DC or RF drives without the risk of breaking the samples. The breakdown at higher temperature has already been reported \cite{Wei1988,Furlan1998, Hohls2002}, and newer aspects related to the effect of localization in Landau levels are discussed in \cite{Aifei2025}.


    \paragraph{\label{sec:VRH} Variable range hopping --}
    To describe the transport at finite energy, we employ the Efros-Shklovskii model introduced 50 years ago~\cite{efros_1975, Shklovskii2024}. It describes the tunneling between localized states in an insulating bulk and extends Mott's original VRH framework \cite{Mott1969} by accounting for long-range Coulomb interactions, which open a soft gap in the density of states (now known as the Coulomb gap). In two-dimensional systems, this leads to a characteristic temperature dependence of the conductivity given by:
    \begin{equation}\label{eg:VRH_temp}
        \sigma(T) = \frac{\sigma_0}{T} \exp\left[-\left(\frac{T_0}{T}\right)^{1/2}\right],
    \end{equation}
    where $k_{\rm B} T_0$ is the hopping energy, set by the localization length $\xi$ and the Coulomb interaction strength: 
    \begin{equation}
        k_{\rm B}T_0=\frac{Ce^2}{4\pi\epsilon_0 \epsilon_r \xi},
    \end{equation}
    where $\epsilon$ is the dielectric constant ($\epsilon_r\simeq3.4$ for hBN) and $C\simeq6.2$ in 2D \cite{Shklovskii2024}. Within this framework, the average hopping distance is $r_{\rm hop} = \xi \sqrt{T_0/T}$, a quantity we will later use to estimate the electrostatic potential $eEr_{\rm hop}$.
    \\
    We extract $T_0$ from temperature-dependent conductance measurements at various filling factors by fitting to Eq.~\ref{eg:VRH_temp}, as shown in Fig.~\ref{fig:results}f). The fits show very good agreement in the two orders of magnitude in temperature accessible with our experimental setup. The extracted fit parameters are shown as a function of the filling factor $\nu$ in Fig.~\ref{fig:T0vsV0}. The activation temperature $T_0$ decreases from the center of each gap $\nu_{\rm c}$ toward its edges. Additionally, we observe that $T_0$ is consistently larger for the cyclotron gaps ($\nu = 6, 10$) than for the spin gaps ($\nu = 4, 8$).
    \paragraph{\label{sec:nonohmic-VRH} Electric field-induced transport --}

    \begin{figure}[t]
        \centering
        \includegraphics[width=\textwidth]{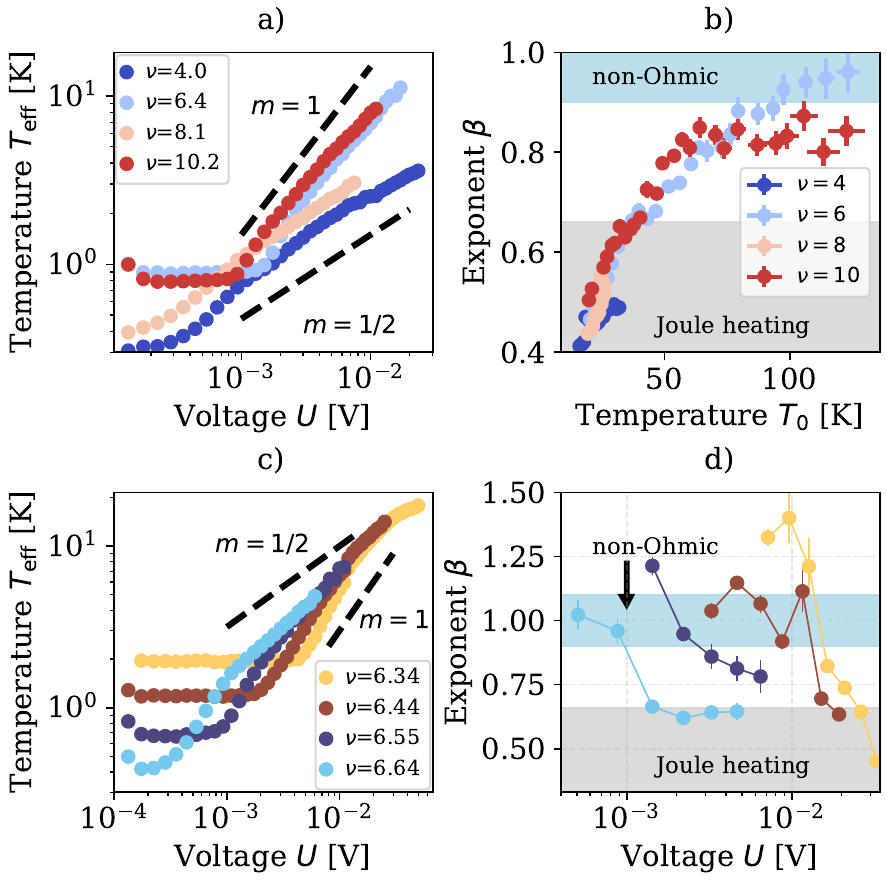}
        \caption{\justifying\textbf{Heating analysis} (a) Effective electronic temperature $T_{\rm eff}$, extracted from the measured conductance using the VRH fit of Fig.~\ref{fig:results}f), as a function of DC bias $U_{\rm DC}$ for four representative gaps. One observes $T_{\rm eff} \propto U$ for $\nu=\num{6.4},\num{10.2}$ and $T_{\rm eff} \propto U^{1/2}$ for $\nu=\num{4.0},\num{8.1}$, as indicated by dashed lines of slope 1 and $1/2$.
        (b) Power-law exponent $\beta$ from $T_{\rm eff} \propto U^{\beta}$ fits, plotted versus the hopping energy $T_0$ at the same $\nu$. Grey and blue regions indicate the expected ranges for Joule-heating–dominated and non-ohmic transport, respectively.
        (c) Same as a) for various filling factors within the $\nu=6$ gap.
        (d) Bias-dependent exponent $\beta$ from sliding-window fits: each fit uses six adjacent data points, with the exponent $\beta$ plotted at the mean $U$ if the MSE is $\textless 10\%$.}

        \label{fig:joule_fit}
    \end{figure} 
   
    In an applied electric field, two complementary mechanisms can govern transport depending on the energy dissipation and thermalization dynamics. In the non-ohmic regime, the applied electric field $E$ modifies the tunneling probability directly while Joule heating is negligible due to an exponentially small conductance. Polyakov and Shklovskii~\cite{Polyakov1993} showed that $E$ acts analogously to temperature, introducing an effective tunneling energy $k_{\rm B} T_{\rm eff}=e\xi E/2 \propto U$. This linear scaling of $T_{\rm eff}$ with bias has been observed experimentally~\cite{Bennaceur2012,Baker2012}, where $T_{\rm eff}$ is the temperature in the Boltzmann distribution, resulting from the field-induced tunneling instead of phonon excitation \cite{Shklovskii1973}. \\
    At higher bias, the conductance is no longer exponentially small and Joule heating becomes significant. In this ohmic regime, electrons are no longer in thermal equilibrium with the lattice due to Joule heating and limited cooling via phonons. The balance between power input $P_{\mathrm{diss}} = \sigma(U) U^2$ and phonon cooling $P_{\text{out}} = \Sigma \left( T_{\rm e}^\alpha - T_p^\alpha \right)$ determines the steady-state electronic temperature:
    \begin{equation}\label{eq:joule}
        T_{\rm e} = \left( T_p^\alpha + \frac{\sigma(U, T_{\rm e}) U^2}{\Sigma} \right)^{1/\alpha},
    \end{equation}
    where $\Sigma$ is a electron-phonon coupling constant and $\alpha$ is for semiconductor typically between 3 and 5~\cite{jezouin2013,lebreton2022,ferguson2025,hill1979}. In the limit of negligible lattice temperature $T_p \rightarrow 0$, this reduces to: $T_{\rm e} \propto  U ^{2/\alpha}$ (with $2/\alpha\in[0.4,0.\overline{6}]$ for $\alpha=3-5$).
    
    These regimes are thus distinguished by their scaling: $T_{\rm eff}\propto U$ in the non-ohmic case, versus $T_{\rm e}\propto U^{2/\alpha}$ in the heating-dominated regime, providing an experimental signature which we investigate in the next paragraph.
    
    \paragraph{\label{sec:crossover} Non-Ohmic and ohmic regime --}
        To identify the dominant transport mechanism, we estimate the effective electronic temperature $T_{\rm eff}$ from our data by inverting the temperature-dependent VRH fits $\sigma(T)$ shown in Fig.~\ref{fig:results}f). 

       The resulting voltage dependence of $T_{\text{eff}}$ is shown in Fig.~\ref{fig:joule_fit}a). Close to the center of the cyclotron gaps ($\nu = 6, 10$), $T_{\text{eff}}$ scales approximately linearly with voltage beyond a certain threshold, consistent with the non-ohmic VRH regime. Below this threshold, the conductance is exponentially suppressed below the measurement resolution. Hence, the effective temperature is mapped onto the onset temperature of conductance. For the smaller spin gaps ($\nu = 4, 8$), we observe a scaling closer to $T_{\text{eff}} \propto U^{1/2}$ at large bias, suggesting Joule heating as the dominant mechanism, while in the center of large gaps (at large $T_0$), non-ohmic VRH prevails.\\
       To analyze this transition regime more systematically, we fit the relation $T_{\rm eff}(U)$ using a power law $T_{\rm eff}\propto U^\beta$. The extracted exponent $\beta$ is plotted as a function of the VRH hopping energy $k_{\rm B}T_0$ in Fig.~\ref{fig:joule_fit}b). $\beta$ increases monotonically from $\simeq0.4$ in small gaps to $\simeq1$ in large gaps, signaling a crossover from Joule-heating transport ($\beta\simeq0.5$) to non-ohmic VRH ($\beta\simeq1$). We further confirm that $\beta$ is frequency independent (see \cite{supplement}), showing that excitation frequency plays little role in graphene. The crossover is explored in detail in the next section.

    \paragraph{Crossover from non-Ohmic to ohmic regimes}

    To further probe the crossover from non-ohmic, field-driven VRH to ohmic transport, we examine the bias dependence of the exponent $\beta$ for intermediate hopping energies $k_{\rm B} T_0$, where our fits yielded values between 0.5 and 1. In Fig.~\ref{fig:joule_fit}c) and ~\ref{fig:joule_fit}d), we focus on filling factors $\nu = 6.34$ - $6.64$, corresponding to $T_0 = 40$ - \SI{180}{\kelvin}. At certain filling factors (e.g. $\nu = 6.34,6.44$), $T_{\rm eff}(U)$ exhibits two distinct slopes, around $\beta=1$ at low $U$ and closer to $\beta=0.5$ at high $U$, indicating a gradual transition between transport mechanisms. The crossover voltage $U_{\rm c}$, defined where the slope changes, increases with $T_0$, consistent with stronger localization requiring larger fields to reach the ohmic regime. 
    
    To quantify this trend, we perform sliding-window fits: six consecutive data points are fitted to $T_{\rm eff}\propto U^\beta$, assigning $\beta$ to the mean $U$ of the window. Despite some scatter due to the small window size (see Fig.~\ref{fig:joule_fit}d)), the results reveal a clear evolution: $\beta$ decreases with bias, signaling a crossover from non-ohmic VRH ($\beta \approx 1$) at low bias to ohmic transport ($\beta \approx 0.5$) at high bias. Moreover, $U_{\rm c}$ increases with $k_{\rm B} T_0$, further supporting that localization strength controls the onset of heating-dominated transport.

\begin{figure}
        \centering
        \includegraphics[width=\linewidth]{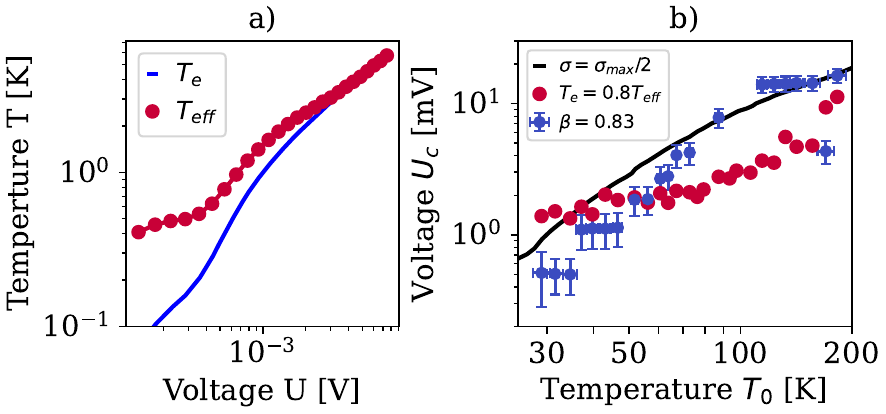}
        \caption{\justifying\textbf{Transition from non-ohmic to ohmic:} 
        (a) $T_{\rm eff}$ estimated from the conductance $\sigma(T)$ and $T_{\rm e}$ estimated from the Joule heating model ($\alpha=3$ and $\Sigma=\SI{1.9e-11}{\watt\per\kelvin^4}$) vs the bias voltage $U$. The curves correspond to a filling factor of $\nu=6.6$.
        (b) Crossover voltage $U_{\rm c}$ vs. hopping energy $T_0$ for $\nu=6$, estimated from where $\beta$ crosses \num{0.83} (blue) or where $T_{\rm e}=0.8\,T_{\rm eff}$ (red). $U_{\rm c}$ increases with $T_0$, indicating that stronger localization requires higher fields. The black line corresponds to the breakdown voltage $U_{\rm BD}$ (where $\sigma=\sigma_{max}/2$).}
        \label{fig:ohmic-transistion}
    \end{figure}

    This crossover can be explained by comparing the gain in electrostatic potential energy $eE r_{\rm hop}$ over a hopping distance $r_{\rm hop}=\xi\sqrt{T_0/T_e}$, and the thermal energy $k_{\rm B} T_{\rm e}$. The non-ohmic transport is more (resp. less) prominent in the regime $eE r_{\rm hop}\gg k_{\rm B} T_{\rm e}$ (resp. $eE r_{\rm hop}\ll k_{\rm B} T_{\rm e}$). As both terms depend on $T_e$ and thus indirectly on $U$, this defines a non-trivial crossover behavior. We focus in the main text on a simple estimate, while a more accurate one is given in \cite{supplement}. We first estimate $T_{\rm e}$ from the Joule-heating model: $T_{\rm e}=(\sigma(U)U^2/\Sigma)^{1/\alpha}$ (assuming here for simplicity $\alpha=3$). The unknown parameter $\Sigma$ is estimated such that the simulated temperature $T_{\rm e}$ aligns with the estimated effective temperature $T_{\rm eff}$ at high values of $U$, where Joule heating dominates and $T_{\rm eff}\simeq T_{\rm e}$ (see Fig.~\ref{fig:ohmic-transistion}a)). We find $\Sigma=0.1-\SI{0.4}{\watt\per\kelvin^4\per\meter^2}$ in agreement with other studies \cite{Fong2012,Fong2013}. 
    The crossover $U_{\rm c}$, defined where $T_{\rm e}\approx0.8\,T_{\rm eff}$, is shown as red dots in Fig.~\ref{fig:ohmic-transistion}b). An alternative estimate, based on $\beta$ dropping below $\beta_c=0.83$, yields similar values (blue dots). Both methods confirm that $U_{\rm c}$ increases with $T_0$, as expected for stronger localization. Furthermore, $U_{\rm c}$ coincides with the conductance onset ($\sigma=\sigma_{\rm max}/2$, black line), consistent with the idea that Joule power $\sigma U^2$ becomes significant at high voltage and large conductance,

    Similar slope changes of $T_{\rm eff}(U)$ were previously reported~\cite{Bennaceur2012}. They were attributed to activated transport into higher Landau levels, at the center of large cyclotron gaps and much larger field ($B=\SI{16.5}{\tesla}$), but heating effects were anticipated for smaller activation energies away from the gap center. By systematically mapping breakdown across a wide range of hopping energies $T_0$, our work shows that the observed slope changes observed in both studies originate from a crossover from non-ohmic, field-driven VRH at large $T_0$ to Joule-heating–dominated transport at smaller $T_0$.

    \paragraph{\label{sec:summary}Summary and conclusions --}
    Our experiments demonstrate that the breakdown of the QH regime in graphene is characterized by a systematic crossover from non-ohmic, field-driven VRH to ohmic, Joule–heating–dominated transport. The crossover voltage $U_{\rm c}$ scales solely with the hopping energy $T_0$, identifying localization length $\xi$ as the key parameter that controls the onset of dissipation. This unified picture explains why cyclotron gaps, with a very small minimal localization length, withstand higher bias fields than smaller gaps, where the localization length saturates due to overlapping Landau levels \cite{Aifei2025}. It also further clarifies the role of heating across a wide range of energy scales.\\
    In contrast to the quantum anomalous Hall effect~\cite{roeper2025}, where frequency-dependent coupling of charge puddles dominates, we find no measurable frequency dependence from DC to \SI{10}{GHz}, showing that breakdown is governed by intrinsic bulk localization physics rather than dynamical drive effects. This work confirms low- and high-frequency breakdown as a sensitive probe of bulk conduction and localization in topological systems and opens a route to systematic studies of dissipation in quantum materials under high-frequency excitation.
    
       


The supporting data and codes for this article are available from Zenodo \cite{zenodo_data}.

\section{Acknowledgements}
    \begin{acknowledgments}
         We warmly thank B. Shklovskii for his insights. This work has been supported by Germany’s Excellence Strategy (Cluster of Excellence Matter and Light for Quantum Computing ML4Q, EXC 2004/1 - 390534769) and the DFG (SFB1238 Control and Dynamics of Quantum Materials, 277146847, projects A04, B07, and C02). It was also funded by the ERC (ERC-2018-STG QUAHQ), by the “Investissements d’Avenir” LabEx PALM (ANR-10-LABX-0039-PALM), and by the Region Ile de France through the DIM QUANTIP. O.M. acknowledges funding from the ANR (ANR-23-CE47-0002 CRAQUANT). K.W. and T.T. acknowledge support from the JSPS KAKENHI (Grant Numbers 21H05233 and 23H02052) , the CREST (JPMJCR24A5), JST and World Premier International Research Center Initiative (WPI), MEXT, Japan.
    \end{acknowledgments}

\bibliography{Literature}
\clearpage
\pagebreak

\appendix
\renewcommand{\thefigure}{S\arabic{figure}}
\renewcommand{\thetable}{S\arabic{table}}
\setcounter{figure}{0}
\setcounter{table}{0}

\onecolumngrid
\section*{Supplementary Material}
\label{sec:appendix}
\twocolumngrid

\subsection{Fitting with generalized VRH law}

The bias dependence of the conductance can be described within a generalized Efros–Shklovskii VRH framework. Independent of whether transport is ohmic or non-ohmic, the current–voltage characteristics follow:
\begin{equation}\label{eq:VRH_field}
    \sigma(U) = \frac{\sigma_0}{U^{2/\alpha}} \exp\!\left[-\left(\frac{U_0}{U}\right)^{1/\alpha}\right],
\end{equation}
where $U_0$ sets the breakdown field and $\alpha = 2/\beta$ reflects the transport regime. Strictly speaking, this form applies to a single regime (either cold or electron regime), but it provides good phenomenological fits even across crossover regions.

\begin{figure}
    \centering
    \includegraphics[width=0.7\linewidth]{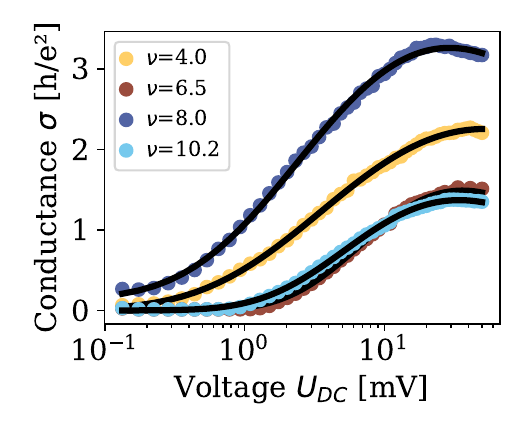}
    \caption{\justifying\textbf{VRH fit of voltage dependence:} Conductance vs. DC bias voltage $U_{\rm DC}$ for different filling factors. The black lines correspond to fits with Eq.~\ref{eq:VRH_field} with $\alpha$, $U_0$, and $\sigma_0$ as fitting parameters (with an offset for not fully developed gaps). The fit results are shown in Tab.~\ref{tab:voltage-vrh-fit}. The fitted exponents confirm that $\nu=6.5,10.2$ are dominated by non-ohmic VRH, while $\nu=4.0,8.0$ are ohmic (Joule-heated). }
    \label{fig:voltage-vrh-fit}
\end{figure}

Figure~\ref{fig:voltage-vrh-fit} shows representative fits for several filling factors, with parameters summarized in Tab.~\ref{tab:voltage-vrh-fit}. 
For $\nu=6.5,10.2$ the fitted $\alpha\approx2$ is consistent with non-ohmic VRH, while for $\nu=4.0,8.0$ we obtain $\alpha\approx4$, indicative of ohmic (Joule-heated) transport. 
This confirms that the VRH description captures both limits and smoothly interpolates between them.

\begin{table}[h]
    \centering
    \begin{tabular}{c|c|c|c}
        \textbf{$\nu$} & \textbf{$U_0$ [\SI{}{\volt}]} & \textbf{$R_K\cdot\sigma_0$ [$\SI{}{\siemens\kelvin^{2/\alpha}}$]} & \textbf{$\alpha$}  \\
        \hline
        4.0\, & \,$0.79\pm 0.06$\, & \,$ 3.7 \pm 0.2$\, & \,$3.9 \pm 0.1$\, \\
        6.5\,& \,$0.20\pm 0.01$\, & \,$ 0.7 \pm 0.1$\, & \,$2.4 \pm 0.1$\, \\
        8.0\, & \,$0.36\pm 0.02$\, & \,$ 3.3 \pm 0.2$\, & \,$3.6 \pm 0.1$\, \\
        10.2\, & \,$0.23\pm 0.01$\, & \,$ 0.8 \pm 0.1$\, & \,$2.7 \pm 0.1$\, \\
    \end{tabular}
    \caption{\justifying\textbf{Fit results of voltage dependence:} The fitted VRH parameter $\sigma_0$, $U_0$ and $\alpha$ that correspond to the curves in Fig.~\ref{fig:voltage-vrh-fit} are shown for different gaps.}
    \label{tab:voltage-vrh-fit}
\end{table}
\subsection{Scaling behavior of normalized fit parameter}
    We next analyze how the VRH parameters scale with the filling factor $\nu$. We compare the characteristic energy scales extracted from the different fits: from thermal activation sweep $k_{\rm B}T_0$, and from the DC bias sweeps ($E_{\rm BD}({\rm DC})$), and RF-driven breakdown ($E_{\rm BD}({\rm RF})$), both defined as $E_{\rm BD}=eU_{\rm BD}$, with the conductance $\sigma(U_{\rm BD})=\max(\sigma)/2$ reaching half of its maximum. 
    
    In this section, all quantities are normalized to their maximum values at the center of the respective gap to account for differences in absolute magnitude.

\begin{figure}[t]
\centering
\includegraphics[width=0.9\textwidth]{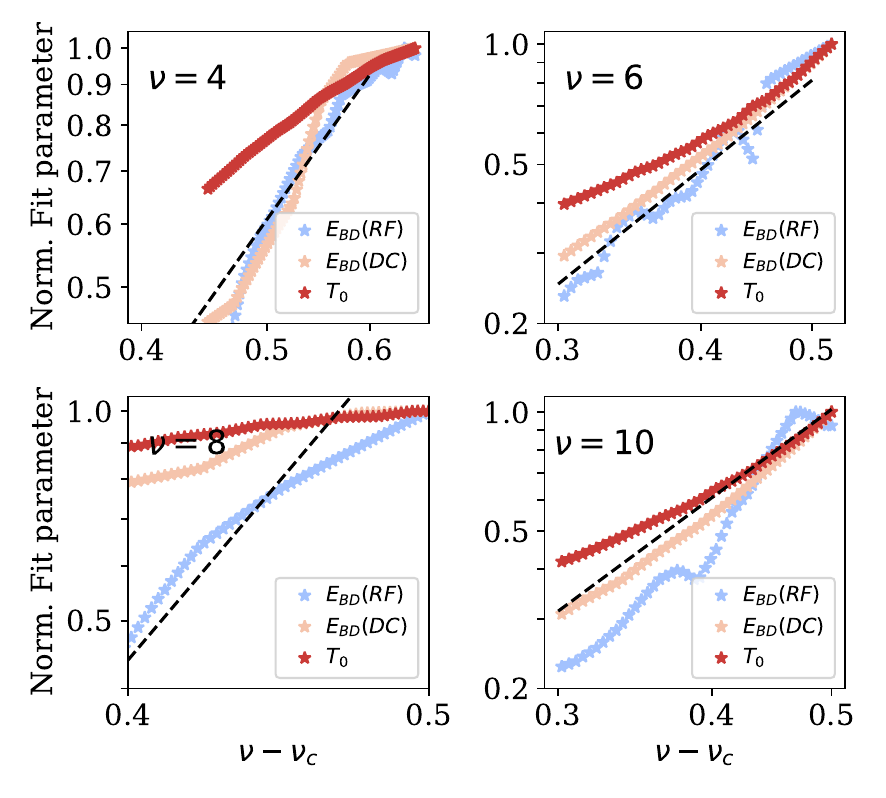}
\caption{\justifying\textbf{Scaling of normalized fit parameters.} Normalized values of $T_0$, $E_{\rm BD}$ (DC), and $E_{\rm BD}$ (RF) as a function of the distance from gap center $\nu - \nu_{\rm c}$ in a log-log plot. The dashed line highlights the theoretical scaling $T_0\propto |\nu-\nu_{\rm c}|^{2.3}$. Cyclotron gaps ($\nu = 6, 10$) follow the expected slope $\gamma=2.3$, while spin gaps ($\nu = 4, 8$) show a smaller slope $\gamma\simeq1$.}
\label{fig:scaling}
\end{figure}

    Figure~\ref{fig:scaling} shows the normalized parameters as a function of the distance from the critical filling factor $\nu-\nu_{\rm c}$, where $\nu_c$ is defined at the conductance peak. The cyclotron gaps ($\nu=6,10$) follow the expected theoretical scaling $T_0 \propto |\nu-\nu_{\rm c}|^{2.3}$, while the spin gaps ($\nu=4,8$) display a weaker slope $\gamma\simeq1$. This deviation likely arises because the spin gaps are less well developed, with multiple Landau levels contributing to hopping transport. These results show that the scaling of transport parameters is strongly affected by the nature of the underlying gap.

\subsection{\label{sec:scaling} Scaling between temperature and field-driven energy scales --}
    \begin{figure}[t]
        \centering
        \includegraphics[width=\textwidth]{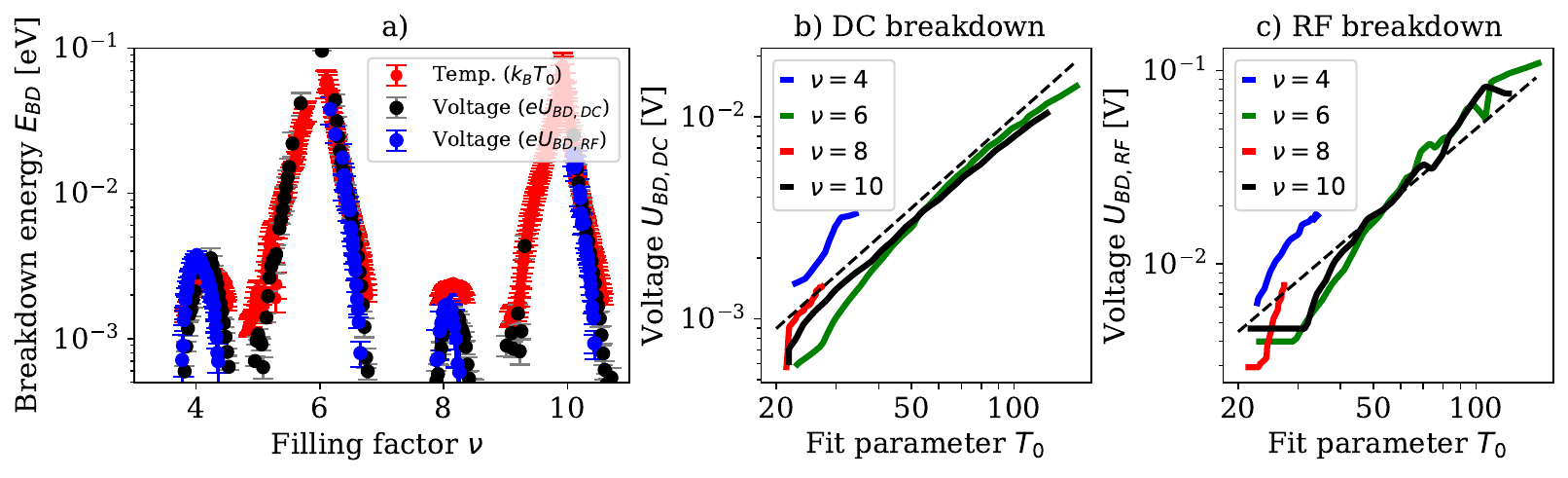}
        \caption{\justifying\textbf{Comparison between thermally and field-driven VRH.}
        (a) Breakdown energy $eU_{\rm BD}/k_{\rm B}$ vs. $T_0$ for DC breakdown. (b) Same for RF breakdown. The log-log slope shows universally a slope of approximately \num{1.5} as shown by the dashed lines with a slope of \num{1.5}, indicating a power law of $U_{\rm BD}\propto T_0^{3/2}$ in DC and RF.}

        \label{fig:T0vsV0}
    \end{figure}  
    Beyond filling-factor dependence, we test the direct relation between the hopping energy $k_{\rm B}T_0$ and the breakdown energies $E_{\rm BD}$. As shown in Fig.~\ref{fig:scaling}, both quantities exhibit similar filling-factor dependence, with maxima at the centers of QH plateaus and minima at transitions. To quantify this behavior, we interpolate $T_0$ and $U_{\rm BD}$ as a function of $\nu$. Plotting $U_{\rm BD}$ versus $T_0$ on a log-log scale (see Fig.~\ref{fig:T0vsV0}a) for DC and b) for RF) reveals a robust power-law relation $U_{\rm BD} \propto T_0^{3/2}$, independent of filling factor or excitation frequency. Notably, spin gaps ($\nu=4,8$) show a slightly steeper slope, but the overall scaling holds. This universal relation highlights, in a different manner, the connection between thermally and field-driven transport in the VRH regime.

\subsection{Frequency dependence}

\begin{figure}[t]
\centering
\includegraphics[width=\linewidth]{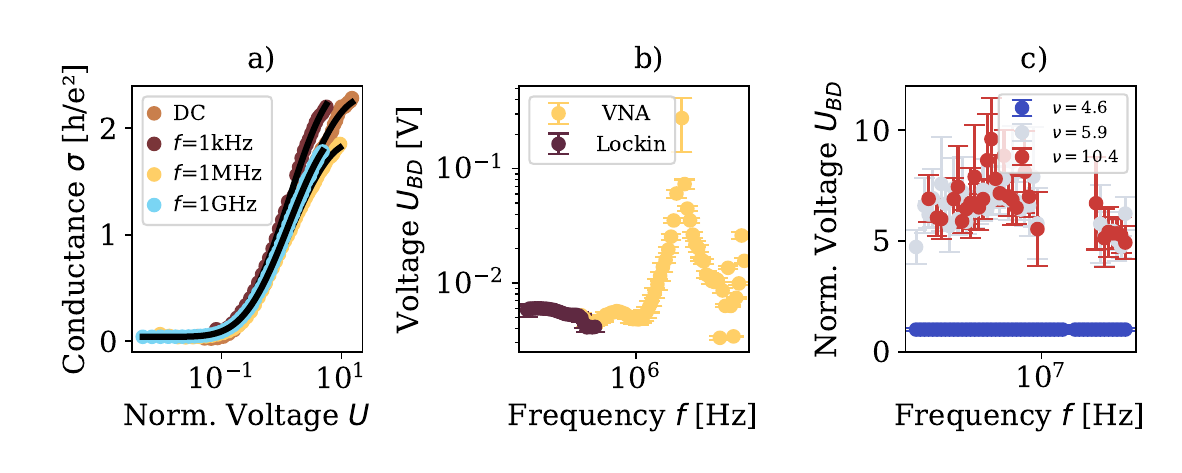}
\caption{\justifying\textbf{Frequency dependence of breakdown at $\nu=4$.} (a) Conductance vs. DC bias $U$ for different excitation frequencies. (b) Extracted breakdown voltage $U_{\rm BD}$ normalized by its value at $\nu=4$, showing weak intrinsic frequency dependence. (c) Control analysis at $\nu=4.6$ demonstrates that observed trends are setup-dependent.}
\label{fig:frequency}
\end{figure}

    We next test whether breakdown depends on excitation frequency. Figure~\ref{fig:frequency}a) shows breakdown curves at $\nu=4$ for various frequencies, which retain the same shape. The extracted breakdown voltage $U_{\rm BD}$ is plotted versus frequency in Fig.~\ref{fig:frequency}b). While $U_{\rm BD}$ appears to increase with frequency, this trend is not intrinsic but due to frequency-dependent transmission losses in the coaxial setup. A resonance feature around \SI{50}{\mega\hertz} is consistent with standing-wave effects from impedance mismatch between sample and lines. These are extrinsic, geometry-dependent artifacts.

    Despite such variations, the underlying breakdown mechanism is frequency independent. The scaling behavior holds over nearly ten decades in frequency, from DC to \SI{10}{GHz}, demonstrating that breakdown is driven by electric-field--induced tunneling between localized states.

\begin{figure}[t]
\centering
\includegraphics[width=\linewidth]{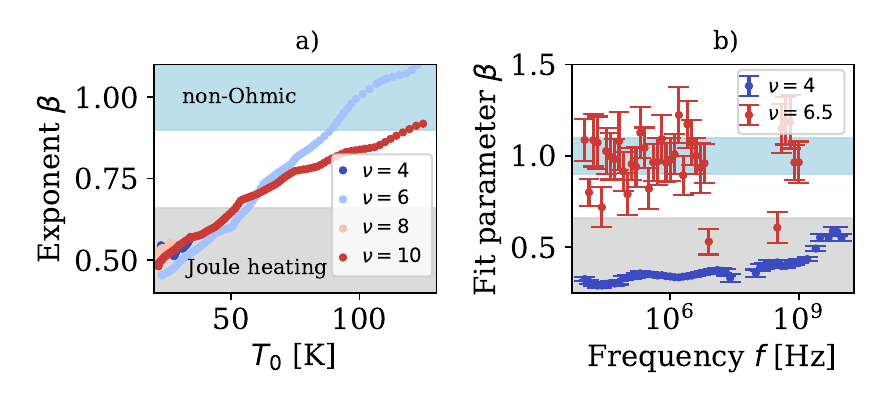}
\caption{\justifying\textbf{Scaling of VRH exponent with activation energy.} (a) Exponent $\beta$ from $T_{\text{eff}}\propto U^{\beta}$ fits vs. $T_0$. Shaded areas mark the expected ranges for Joule heating (gray) and non-ohmic VRH (blue). (b) Frequency dependence of $\beta$ for $\nu=4$ and $\nu=6.5$, corresponding to ohmic- and non-ohmic--dominated regimes, respectively.}
\label{fig:beta_vs_t0_VRH_fit}
\end{figure}
\subsection{Extended data for bias dependence}
    
\begin{figure*}[t]
            \centering
            \includegraphics[width=\textwidth]{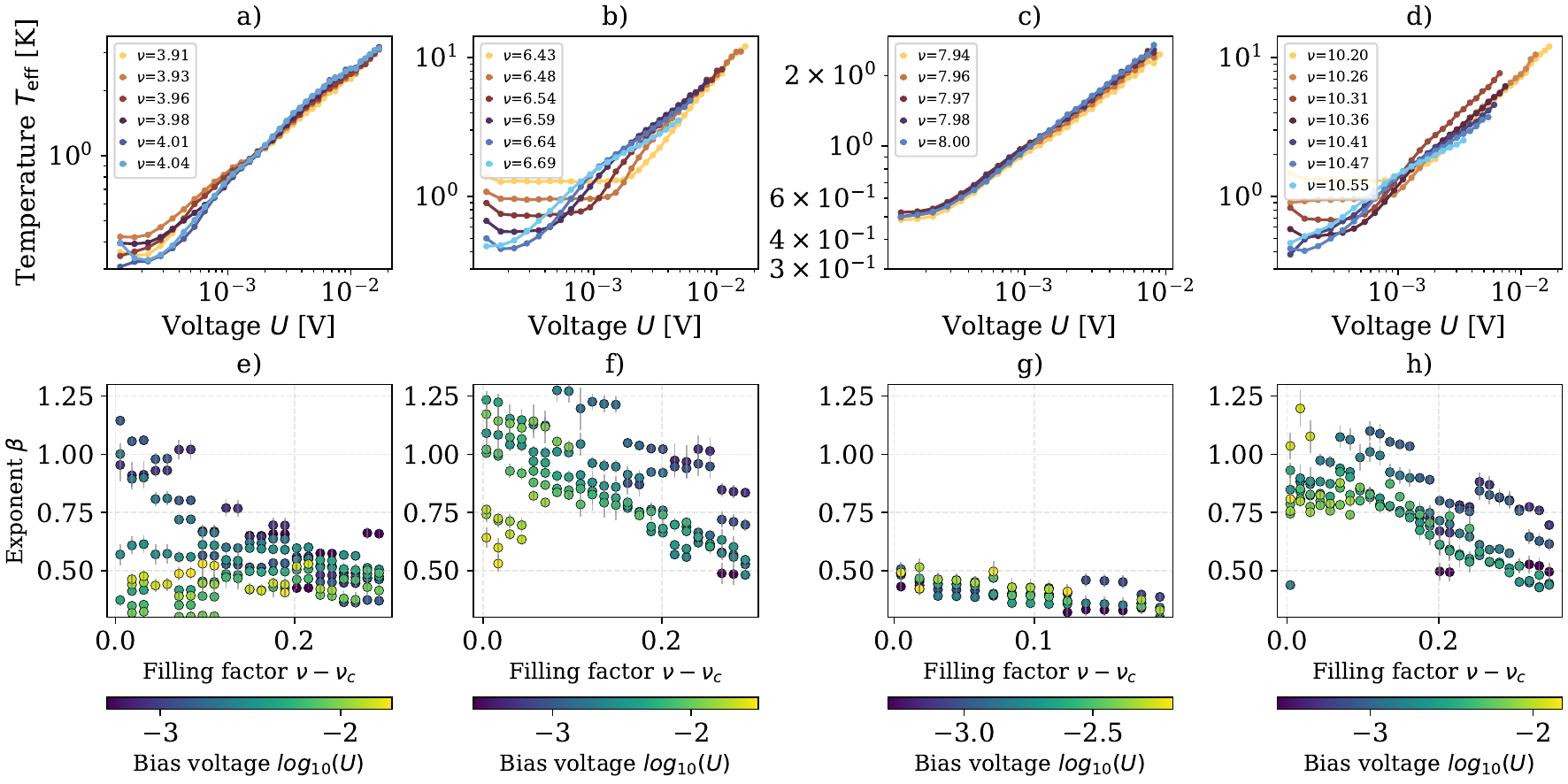}
            \caption{\justifying\textbf{Extended data for bias dependence}: a)-d): Effective Temperature $T_{\rm eff}$ vs. bias voltage $U$ for filling factor around $\nu=4,6,8,10$. The slope changes from 1, for small bias and high activation energy to 0.5 for large bias and small activation energy. e)-h) Analysis of the power law exponent $\beta$ as a function of the filling factor (shown on x-axis) and bias voltage (shown as color), for a sliding window: a power law fit for 6 adjacent voltages with the mean voltage of each window shown as color. It shows a decreasing exponent with increasing filling factor and bias voltage.}
    
            \label{fig:bias_dependence}
        
    \end{figure*}

    In the main text, we demonstrated that the effective electronic temperature $T_{\rm eff}$ follows a power-law scaling with applied bias and a crossover from non-ohmic to ohmic conduction with increasing bias. Here, we provide extended data around all four gaps ($\nu=4,6,8,10$).
    Figures~\ref{fig:bias_dependence}a)–d) show $T_{\rm eff}$ as a function of $U$, highlighting a crossover from linear scaling ($\beta\simeq1$) to square-root scaling ($\beta\simeq1/2$). 
    To quantify the evolution of $\beta$, we perform sliding-window fits of the form $T_{\rm eff}\propto U^\beta$ using six adjacent bias points. 
    The resulting exponents are plotted in Figs.~\ref{fig:bias_dependence}e)–h), revealing a systematic decrease of $\beta$ with both increasing filling factor ($x$-axis) and bias voltage  (color). This analysis confirms that the transport continuously evolves from non-ohmic to Joule-heating–dominated conduction with increasing bias voltage, with a threshold that increases with $k_{\rm B}T_0$.
    
\subsection{Estimates of the crossover amplitude}

\begin{figure}
    \centering
    \includegraphics[width=0.9\linewidth]{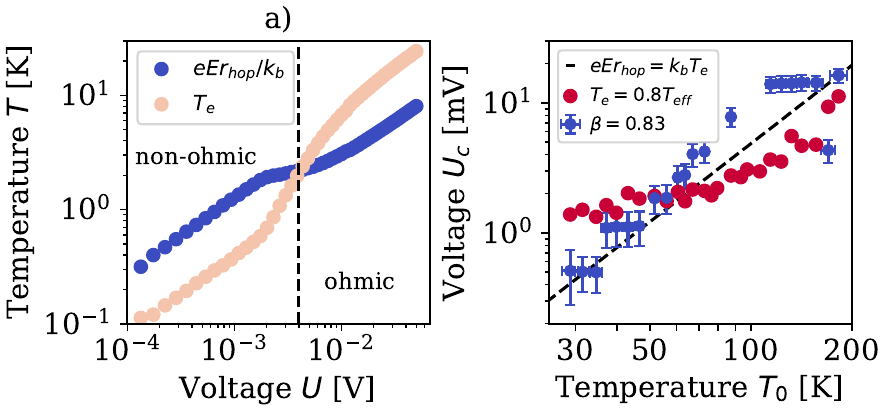}
    \caption{\justifying\textbf{Estimate of crossover amplitude:} (a)
    Electric field energy $eEr_{\rm hop}$ and estimated electron temperature $k_BT_e$ vs. bias $U$ for $\nu=6.9$ (The electric
    field energy has a numerical pre-factor of 0.1). Their
    intersection marks the onset of ohmic conduction. $T_e$
    is estimated via Joule heating with $\alpha=3$ and $\Sigma=\SI{8e-12}{\watt\per\kelvin^3}$. (b) Crossover voltage $U_{\rm c}$ vs. hopping energy $T_0$ for $\nu=6$, estimated from where $\beta$ crosses \num{0.83} (blue) or where $T_{\rm e}=0.8\,T_{\rm eff}$ (red). The black line corresponds to the prediction transition where $eEr_{\rm hop}=k_BT_e$ for varying $T_0$.}
    \label{fig:eEr_vs_kT}
\end{figure}
A simple estimate of the crossover can be obtained by comparing the electrostatic potential energy over the hopping distance, $eE r_{\rm hop}$, with the electronic temperature, $k_{\rm B}T_{e}$. Within the Efros--Shklovskii framework, the hopping length is $r_{\rm hop}=\xi\sqrt{T_0/T_e}$, while the electric field is approximated as $E=U/w$ with $w\simeq\SI{2}{\micro\meter}$ the contact separation. For the electronic temperature we use the Joule-heating model, $T_e=(\sigma(U)U^2/\Sigma)^{1/\alpha}$, assuming $\alpha=3$. The coupling constant $\Sigma$ is calibrated such that $T_e$ coincides with the experimentally extracted effective temperature $T_{\rm eff}$ at high bias, where Joule heating dominates (see Fig.~\ref{fig:ohmic-transistion}a).

Using this procedure, we find that $eE r_{\rm hop}$ is smaller than $k_{\rm B}T_e$. A reasonable agreement with experiment is obtained if a numerical prefactor of order $0.1$ is included in the field energy, which likely reflects uncertainties in system parameters such as the effective field distribution or the precise definition of $r_{\rm hop}$. Fig.~\ref{fig:eEr_vs_kT}a) shows $eE r_{\rm hop}$ and $k_{\rm B}T_e$ as a function of the voltage $U$. We observe a crossing due to an onset of $T_e$, which can be understood from the onset of bulk conductance, which leads to an onset of Joule power $P\propto \sigma U^2$. The model predicts a crossover voltage $U_c$ that increases systematically with $T_0$, in good agreement with the experimental data (Fig.~\ref{fig:ohmic-transistion}b). The crossover voltage $U_c$ is estimated by calculating the shift in $U_c$, while fixing all parameter but $T_0$.

\subsection{Robustness against magnetic field and carrier type}
\begin{figure}[t]
\centering
\includegraphics[width=\linewidth]{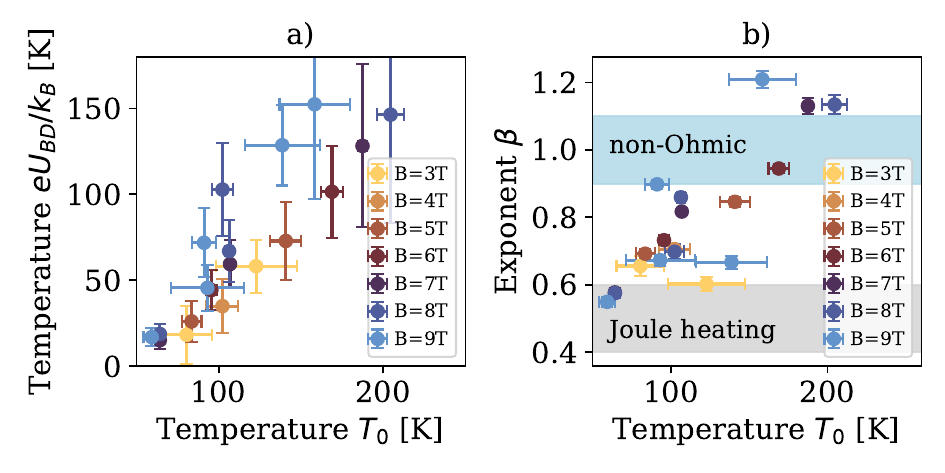}
\caption{\justifying\textbf{Magnetic field dependence.} (a) Breakdown energy $eU_{\rm BD}/k_{\rm B}$ vs. $T_0$ for the $\nu=6$ gap at different magnetic fields, showing field-independent scaling. (b) Power-law exponent $\beta$ vs. $T_0$, showing transition from ohmic to non-ohmic transport with increasing hopping energy.}
\label{fig:field_dependence}
\end{figure}

Finally, we analyze the robustness of the breakdown mechanism against variations in carrier type and magnetic field. Figure~\ref{fig:field_dependence} shows that for $\nu=6$ the relation between $U_{\rm BD}$ and $T_0$ is unchanged across different magnetic fields, confirming that the relevant energy scales are set by localization rather than absolute field strength.

\begin{figure}[t]
\centering
\includegraphics[width=\linewidth]{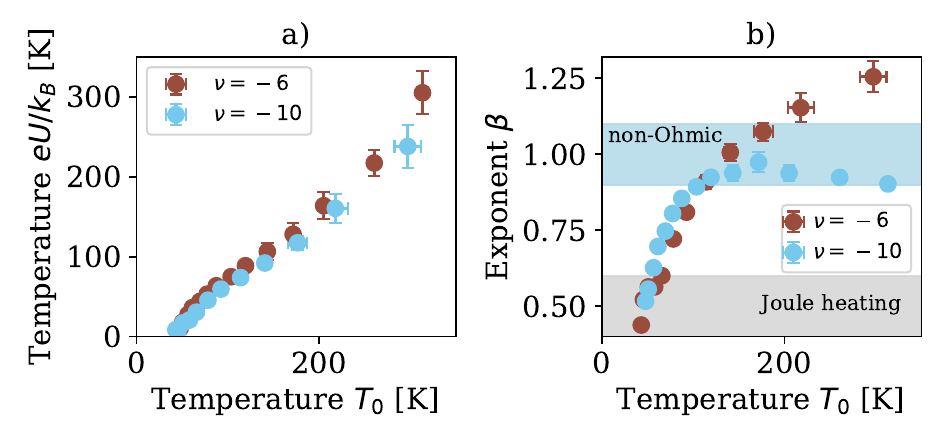}
\caption{\justifying\textbf{Hole doping.} (a) Breakdown energy $eU_{\rm BD}/k_{\rm B}$ vs. $T_0$ for hole-doped states $\nu=-6,-10$ at $B=\SI{9}{\tesla}$, showing the same breakdown mechanism as for electrons. (b) $T_0$-dependence of exponent $\beta$, confirming the same crossover from ohmic to non-ohmic transport.}
\label{fig:VRH_hole_doping}
\end{figure}

We also investigate hole-doped states at $\nu=-6$ and $\nu=-10$ (Fig.~\ref{fig:VRH_hole_doping}). The scaling of $U_{\rm BD}$ with $T_0$, as well as the evolution of $\beta$, closely mirrors the behavior on the electron side. This symmetry demonstrates that the VRH-based breakdown mechanism applies universally, independent of carrier type.

\newpage

\end{document}